\documentclass[twocolumn,showpacs,preprintnumbers,amsmath,amssymb,showkeys,nofootinbib]{revtex4}

\usepackage{graphicx}
\usepackage{dcolumn}
\usepackage{bm}

\begin{document}



\title{Self-organized criticality in quantum gravity}

\author{Mohammad H. Ansari}
\email{mansari@perimeterinstitute.ca}
\author{Lee Smolin}%
\email{lsmolin@perimeterinstitute.ca}
\affiliation{University of Waterloo, Waterloo, On, Canada N2L 3G1 }%
\affiliation{Perimeter Institute, Waterloo, On, Canada N2L 2Y5}%

\date{\today}

\begin{abstract}

We study a simple model of spin network evolution motivated by the
hypothesis that the emergence of classical space-time from a
discrete microscopic dynamics may be a self-organized critical
process. Self organized critical systems are statistical systems
that naturally evolve without fine tuning to critical states in
which correlation functions are scale invariant.  We study several
rules for evolution of {\it frozen spin networks} in which the
spins labelling the edges evolve on a fixed graph. We find
evidence for a set of rules which behaves analogously to sand pile
models in which a critical state emerges without fine tuning, in
which some correlation functions become scale invariant.

\end{abstract}

\pacs{04.60.Pp, 04.60.-m, 05.65.+b, 89.75.Da}

\keywords{dynamics; evolution; self-organized criticality; complex
system; quantum gravity; spin network}

\maketitle

\section{Introduction}

Since the work of Wilson and others \cite{Wilson} it has been
understood that the existence of a quantum field theory requires a
critical phenomena, so that there are strong correlations on
scales of the Compton wavelength of the lightest particle. If this
scale is to remain fixed as the ultraviolet cutoff length is taken
to zero, the couplings must be tuned to a critical point, so that
the ratio of the cutoff to the scale of the physical correlation
length diverges. This requires asymptotic scale invariance of the
kind found in second order phase transitions.

Similar considerations apply to quantum gravity in a background
independent formulation such as loop quantum gravity, or causal
set models. The problem is not alleviated if the theory is shown
to be finite due to there being a physical ultraviolet cutoff, as
in loop quantum gravity\cite{LQG}. Instead,  the need for a critical
phenomena is even more serious as there is no background geometry.
This means that away from a critical point the system may not have
any phenomena that can be characterized by scales much longer than the Planck
length.   That is to say, the volume, measured for example, by the number of events,
may become large, but there may still be no pairs of events further than a few Planck times
or lengths
from each other. This is
seen in detail in models whose critical phenomena has been well
studied, such as dynamical triangulation models\cite{DT} and Regge
calculus\cite{Regge}. Away from possible critical points, the average distance
between two nodes or points need not grow as the number of events
(or the total space-time volume) grows.  Instead, one sees that for typical couplings,
statistical measures of the dimension, such as the hausdorff dimension,
can go to infinity or zero.

In  equilibrium statistical mechanical systems, critical
phenomena of the kind required for a system to exhibit a large hierarchy of scales
is only found on
renormalization group trajectories that flow to ultraviolet fixed
points of the renormalization group.  It typically requires a
 fine tuning of many couplings to put the system on a physical
renormalization group trajectory. This may be seen to be generally
problematic when the system under study is not a laboratory
experiment, but is conjectured to be a theory that is both
fundamental and cosmological, for in this case who is to do the
fine tuning required for our macroscopic world to emerge?

Given this, it is very interesting that there are
some systems whose
parameters reach critical behavior, with scale invariant correlations,
without any necessity to tune the parameters from outside
the system. These are typically non-equilibrium systems, which
reach critical behavior after evolution in real time from a
random  start.  Among these are
self-organized critical systems studied by \cite{bak, bak2,
turcotte}.

It is then attractive to consider the idea that the critical
behavior necessary if classical space-time is to emerge from a
background independent quantum theory of gravity arises from a
process analogous to self-organized critical phenomena.  This idea
was  proposed earlier\cite{QG_SOC}, where it was proposed that
the low energy limit of quantum gravity might be analogous to a system
evolving to a self-organized critical behavior such as directed percolation.
This idea was then studied in some detail by
Borissov and Gupta \cite{borissov} in the case of a certain
simplification of loop quantum gravity.   In this simplification,
the graph on which a spin network basis state is defined does not
evolve, rather the spin labels evolve on a fixed graph.  Such
rules define a class of theories we call {\it frozen spin network
theories.} Moreover, the identities that impose gauge invariance
at vertices are not imposed as conditions on states, instead the
dynamics is chosen so that the system evolves to gauge invariant
states.

Borissov and Gupta in \cite{borissov} did not find  a set of evolution rules which are self-organized critical.
Here we study a new set of evolution rules, and show evidence that at least one of them is self-organized
critical.

Before going on we note that our results have one severe
limitation.  We work here not with quantum gravity per se,
but with the classical statistical mechanics of spin networks. The
evolution rules we study are stochastic rather than quantum
mechanical; they are described by real probabilities rather than
complex amplitudes.  Whether the considerations of self-organized
critical systems can apply to critical phenomena in quantum
systems is presently unknown, to our knowledge there is as yet no
example of quantum self-organized critical phenomena.  Nor can we
naively apply the method of Euclidean continuation as is done in
conventional quantum field theory, by means of which quantum
amplitudes are related to statistical mechanical problems. The
reason is that in quantum gravity there is no preferred time
coordinate by means of which the Euclideanization can be done.

To make this paper self contained for interested readers in both
quantum gravity and statistical physics, we give very brief
introductions to spin network states in section 2 \footnote{Those
readers wanting a more detailed introduction to loop quantum
gravity are encouraged to look at \cite{LQG}.} and to
self-organized criticality  in section 3. In section 4 we suggest
two different classes of propagation rules for frozen spin
networks. The first class is based on choosing an edge at random
and changing its color by a constant value and then making the
network gauge invariant. In this class we generalize a model which
was studied for one special propagation rule in \cite{borissov}.
Some of the rules in this class were studied and it was seen that
none of them exhibited self-organized criticality. The second
class is based on choosing a vertex at random among all vertices
of a trivalent spin network and changing the colors of its three
incident edges by a constant even value.  We do find a rule in
this class that exhibits power law behavior, which is suggestive
of self-organized critical systems.

Section 5 presents some of the results of a numerical study we carried out
which provides evidence that members of this second class of rules
 exhibit self-organized criticality. This is followed by our conclusions.

\section {Spin network states}

For the purposes of this paper a spin network is a combinatorial
labelled graph. It consists of a graph $\gamma$,
 which consists of a finite number of oriented edges $e_{1}, e_{2},
\cdots $ incident at vertices $v_{1}, v_{2}, \cdots $. The edges are labelled by the irreducible representation of
a Lie group, $G$. In the case of  canonical quantum gravity in $3+1$ dimensions, $G=SU(2)$,
so that the labels on edges are spins. The
\textit{color} of an edge is defined as twice the spin, $c_{i} = 2 j_{i}$.

In this paper we will consider only trivalent spin networks which
may be embedded in a plane. Dual to such an imbedded trivalent
spin network is a triangulation of a region of
space\cite{fotini-dual}. The length of a side in the triangulated
network is proportional to the color of its dual edge in the  spin
network, $ 2 l_{side} = l_{Planck} \cdot c_{edge}$ \cite{rovelli}.
The triangle inequalities hence provide constraints on the lengths
of the sides of a triangle. Therefore there is a constraint on the
colors of incident edges at a vertex.  The constraint is called
{\it the gauge-invariance constraint} because it also corresponds
to the spin network states being gauge invariant, in the sense
that they are solutions to Gauss's law\cite{LQG}\footnote{In the
case of nodes with valence higher than three, the implication of
Gauss's law is more complicated. Each vertex $v_{i}$ of the spin
network is labelled by an invariant, in the product of the
representations of the edges incident on it. }. It can be shown
that the constraint on a vertex is:
\begin{eqnarray}
\label{triangles} a + b \geq c,\ \ a + c \geq b,\ \  b + c \geq a;\\
\label{even}  a + b + c = even.
\end{eqnarray}

where $a$ ,$b$ and $c$ are the positive integer colors of three
incident edges at a vertex.

In loop quantum gravity, spin network states evolve by the application of local
evolution rules, which apply to a single node or a small number of neighboring
nodes\cite{LQG}.  In the dual picture, these involve a small number of neighboring triangles
\cite{fotini-dual}.  The evolution rules have been derived in both a hamiltonian
and path integral framework and come in several versions.  Here we study a class
of rules which are greatly simplified from those studied in the literature. We keep one
key feature of the rules derived by quantization of the classical theory, which is that
they involve the modification of a spin network state by the addition or subtraction of a small
loop of non-abelian flux. This corresponds to the fact that the Einstein equations are
first order in the curvature of the spacetime connection. The addition or subtraction of
a loop of electric flux corresponds precisely to the multiplication of the state by a
small Wilson loop of the spacetime connection.

In the exact forms of dynamics of LQG, the loop of spacetime
connection is multiplied by operators in the spacetime metric,
which have the effect of gluing the loop to the graph representing
a state in a way that preserves gauge invariance (represented in
the dual picture by the triangle inequalities.)  Thus, the effect
of the dynamics is to evolve the graphs from gauge invariant
configurations to other gauge invariant configurations which
differ by the addition or subtraction of a loop of flux.

Here we propose a two step dynamical process which has the same
effect. The first step is to simply add or subtract loops of flux.
As we will see, this can result in a state in which the triangle
inequalities are not satisfied. The second step is to adjust the
labels on nearby triangles so as to ensure that the result
satisfies the triangle inequalities. Thus, gauge invariance is
achieved only in the end; it comes as a result of a relaxation
process which involves the addition or subtraction of more loops
of flux.  This gives rise to avalanches of moves, whose statistics
gives rise to scale invariant dynamics.

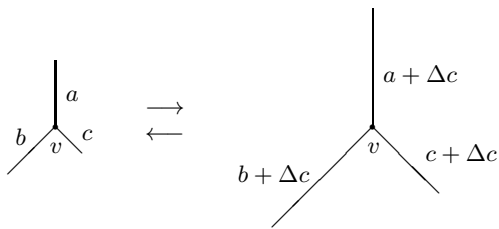
\begin{figure}[h]
\begin{picture}(200,80)
\put(30,40){\line(0,1){25}} \put(30,40){\line(1,-1){10}}
\put(30,40){\line(-1,-1){18}} \put(34,50){$a$} \put(15,33){$b$}
\put(40,35){$c$} \put(30,40){\circle*{2}} \put(28,30){$v$}
\put(64,45){$\longrightarrow$} \put(64,35){$\longleftarrow$}
\put(150,40){\line(0,1){45}} \put(150,40){\line(1,-1){25}}
\put(150,40){\line(-1,-1){38}} \put(154,57){$a+\Delta c$}
\put(99,19){$b+\Delta c$} \put(170,27){$c+\Delta c$}
\put(150,40){\circle*{2}} \put(148,30){$v$}
\end{picture}
\caption{The dynamics of quantum gravity is represented by the
addition or subtraction of loops of flux, corresponding to the
fact that the Einstein equations are linear in the curvature of
the spacetime connection. $a$, $b$ and $c$ are the colors of the
the edges incident at the vertex $v$ and $\Delta c$ is a positive
or negative integer.}
\end{figure}

\section{Self-organized criticality}

A self-organized critical (SOC) system is one which has critical,
that is scale invariant behavior, without fine tuning of
parameters. The earliest example of such a system is the sandpile
of Bak, Tang and Wiesenfeld\cite{bak}. Since then, many such
systems have been studied, including models of phenomena as
diverse as biological evolution, earthquakes, astrophysical
phenomena and economics\cite{bak2}.

One way that SOC systems are identified is by measuring the distribution in space and time
of events in the system's evolution, and looking for power law behavior.
A set of events which is contiguous in both
time and space is called an {\it avalanche}.
 Self-organized criticality (SOC) occurs when the
distribution of the sizes of avalanche follows a power law\cite{bak,bak2,turcotte}:
\begin{equation}
\label{powerlaw}
P(s) = s^{-\alpha}
\end{equation}
where $\alpha$, $s$ and $P(s)$ are a positive constant, the size
of avalanche, and the distribution of a size of avalanche,
consecutively. Because the distribution is power law rather than
exponential, there is no preferred scale that characterizes the
avalanches. There is no largest avalanche, and no typical size for
an avalanche. Hence we can say that the system exhibits the same
structure over all scales.

\section{Evolution rules for frozen spin networks}

Our aim is to find evolution rules that realize SOC in 2d planar trivalent spin networks.  For
simplicity we consider frozen spin networks, which are ones for
which the labels change but the underlying graph remains fixed\footnote{There is a limited
sense in which the topology of our graphs can change, which is when edges have length
zero.  When one edge of a triangle is zero, gauge invariance requires the other two edges
are equal. The triangle then can be considered to have disappeared.}. In
this context we may still attempt to mimic the basic features of
the Hamiltonian constraint of quantum gravity such as the fact
that the dynamics consists of terms which add elementary loops of
flux to the original graph.

We begin by constructing a fixed triangulated triangle graph on
which we will define and study evolution of the labels (Figure 2).

\begin{figure}[h]
\begin{center}
\includegraphics[width=90pt]{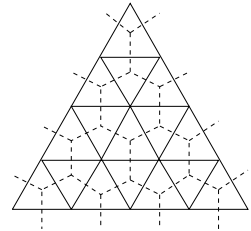}
\caption{Frozen spin network (dashed network) and its dual, the
triangulated triangle (solid lines). }
\end{center}
\end{figure}

We then construct the dual spin network by connecting the centers of each triangle to the centers of the adjacent
triangles.   The result is a trivalent spin network, with boundary given by the dual of the segments of the edges
of the original triangle.

The evolution rules we will invent are designed to be analogous to
the rules by which a sandpile model evolves. First, sand is
randomly dropped onto the pile. The pile is in equilibrium so long
as  the slope of the pile is not too much.  If a new piece of sand
causes the slope to exceed that value, the sand flows, till a new
equilibrium is established. Thus the evolution rules have four
steps: 1) drop sand randomly, 2) check to see if the slopes are
too much, 3) if so move sand locally until all slopes are reduced
below the condition for equilibrium. 4) Go back to step one.

We may consider an analogous set of rules for colors to evolve on
a graph. The hight of the pile is analogous to the color. The
condition that the slope be not too much will be replaced by the
condition that gauge invariance is preserved at each node.  Thus,
the rules we will consider also have four steps: 1) add or
subtract colors to randomly chosen edges. 2) check to see if the
gauge invariance condition is satisfied at all nodes. 3) if it is
not, then move the colors at edges adjacent to non-gauge invariant
nodes around, till gauge invariance is restored. 4) go back to
step one.

The process by which sand redistributes itself on the pile till
equilibrium is re-established is called an {\it avalanche}.  The
size of an avalanche is the number of moves it takes to restore
stability.  When a pile has reached a self-organized critical
state many slopes are just at the value below that which causes
sand to flow downwards. Once this state is reached the
distribution of sizes of avalanches becomes scale invariant.

By analogy, the process by which colors re-arrange themselves on the graph may also be
called an avalanche. If the network reaches a critical state, many nodes will be in a state
where one more addition of a color causes gauge invariance to be satisfied.  This means
that the dual triangle is degenerate, so that the triangle inequality is just barely satisfied.
We seek rules such that, once a sufficient number of nodes are in such a critical state,
the distribution of sizes of avalanches become scale invariant.

We may consider rules in which the
 color added  $\triangle c$ to a vertex is always even
or always odd. The difference between them is as follows. Adding
an odd color to an edge, say $\triangle c = \pm1$, will always
cause the gauge invariance condition (\ref{even}) to be violated.
But if we add (or subtract) even colors, the situation is more
complicated, as gauge invariance at the adjacent nodes will
sometimes be satisfied and sometimes be violated. This is
analogous to the case of a sandpile in which a new piece of sand
sometimes will and sometimes will not increase the slope to a
value where an avalanche starts.  We found by numerical simulation
that critical phenomena can occur in the latter case in which the
changes are even.  The case in which the changes are odd does not
seem to evolve analogously to a sandpile model.

In the case that three incident edges at a vertex have the
particular colors which make one of the three conditions of
(\ref{triangles}) saturated we call the vertex a {\it flat
vertex}. If the initial edge that has accepted $\triangle c$ is
the one with the largest color, adding  $\triangle c = +2$ to it
will violate gauge invariance. If the initial edge has the
smallest color, subtracting  $\triangle c = -2$ from it will
violate gauge invariance. A vertex such as this, where gauge
invariance is violated (or the triangle relation fails for the
dual triangle), will be called a {\it GNI}, for gauge,
non-invariant vertex.

Models of this type fall into two classes according to whether the
random changes are made at edges or nodes. In the first class, we
choose \emph{an edge} at random and change its color by an adding
or subtracting $\Delta c$. In the second class  we choose \emph{a
vertex} at random and change the colors of all the three edges of
it by adding or subtracting $\Delta c$. We can think of the latter
case as one in which closed loops of dual flux are added around
each node, in rough analogy with the evolution rules in loop
quantum gravity and spin foam models.

\subsection {Random edge models}
The evolution of spin network\footnote{The spin network can be a
planar or a closed network. By  a closed spin network we mean a
network in which there is no boundary in that and by walking on
edges we will return to the initial point. For example a
tetrahedron with 4 vertices and 6 edges can be thought of as a
simple closed spin network.} can be defined as changing the color
of one edge, chosen at random, by $\triangle{c}=\pm 2 $ and
checking the conditions (\ref{triangles},\ref{even}) at all
vertices. The propagation rule for recovering the possible
violation of the gauge conditions can be defined in different
ways. The following are examples  of possible  propagation rules
for this class of models.
\newcounter{bean}
\begin{list}
{\Roman{bean}.}{\usecounter{bean}
\setlength{\rightmargin}{\leftmargin}} \item In the case that
adding  color +2 to the initial edge produces a GNI-vertex at its
ends, the propagation rule on the vertex can be chosen to be:

\begin{itemize}
\item adding $\triangle c = +2$ to one of the two less-colored
edges, or \item adding $\triangle c = +2$ to both of the
less-colored edges, or \item passing the added color +2 from the
initial edge to one of the two less-colored edges.
\end{itemize}

\item In the case that subtracting color -2 from the initial edge
produces a GNI-vertex, the propagation rule can be chosen to be:

\begin{itemize}
\item subtracting $\triangle c = -2$ from the largest edges, or
\item adding $\triangle c = +2$ to one of the less-colored edges,
or \item adding $\triangle c = +2$ to both of the less
colored-edges, or \item passing the added color -2 from the
initial edge to one of the other two edges.
\end{itemize}
\end{list}

In either case, we construct a model in which we:

\begin{itemize}
\item initialize a 2d spin network with random but gauge-invariant
colors on its edges. \item choose an edge at random. and change
its color by adding (subtracting) to (from) a constant value of
color $\triangle c $. \item check gauge invariance condition
(\ref{triangles},\ref{even}) from the very first vertex to the
last vertex. \item propagate $\triangle c $ from a GNI-vertex to
other vertices by a propagation rule until the whole network
becomes gauge invariant again. \item store the number of updated
edges as the size of avalanche. \item repeat these steps a large
number of times in order to see the behavior of spin network in a
long time.
\end{itemize}

Borissov and Gupta \cite{borissov} studied a particular
propagation rule on a 2d planar spin network. The model has one
parameter, which is a probability, $p$.  The evolution rule is as
follows. A vertex is chosen randomly. The  edge with  biggest
color among the three incident edges at a vertex  is then evolved.
The color of that edge is increased by  +2  with probability $p$
or decreased, by  -2, with probability $1-p$. Similarly, if the
arbitrary edge is the smallest one the color -2 is subtracted from
the edge with probability $p$ or the color +2 is added to that
with probability $1-p$.

They report that the rule, for different values of $p$, produces
an exponential distribution of avalanches $P(s)\sim
e^{-s/\sigma}$, where $\sigma$ is the decay constant.\footnote{
They report that $\sigma$ reaches a maximum value when the
probability $p$ becomes close to 0.4.} This evolution does not
exhibit SOC on a 2d planar spin network. This means the process of
the recovery of gauge invariance under the special propagation
rule they have proposed, is not self-similar.

We have studied some of the above propagation rules for
open 2 dimensional planar spin networks.  In none of these cases did the distribution of avalanche show scale invariant behavior.

We also studied various rules for colors evolving on closed
graphs including a tetrahedron and a network like a Bucky ball
with 60 vertices and 90 edges. We  did not see evidence of scale
invariant behavior for any closed network we studied. We suspect
that a closed system is less likely to exhibit scale invariant
behavior because the  flux cannot leave the system. (For a good
review of the  role of boundary in sandpile model refer to
\cite{carreras}).  An SOC system is typically an open system, with
a flow of energy or matter through it. It is the flow of energy or
matter through the system that drives the self-organization of the
system.

In order to understand  these models,  it is  useful to
consider the graphical representation shown in
Figure 3.  In the figure, we associate each of three axes with the color on an
edge incident to a given node.

\begin{figure}[h]
\includegraphics[width=150pt]{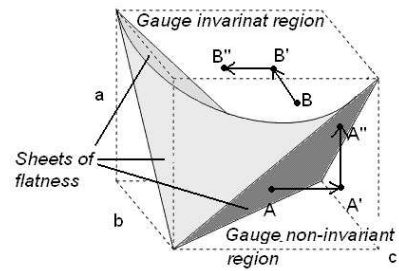}
\caption{\textit{3d color space of each vertex.} All color-points
inside the pyramid satisfy the conditions (\ref{triangles},
\ref{even}). Point $A$ represents a flat vertex whose evolution
kicks it out of the gauge invariant pyramid ($A'$) and using a
propagation rule can make it flat again. ($A''$)}
\end{figure}

The triangle inequalities (\ref{triangles}) divide the 3d color space into two different
regions. All gauge-invariant vertices are located inside or on the boundary of a pyramid
bounded by three surfaces, which are given by the equations,
\begin{equation}
\label{sheets} a + b = c,\ \ a + c = b,\ \  b + c = a;
\end{equation}
where $a$, $b$ and $c$ are the colors of the three incident edges
on a vertex. We call the three boundaries \textit{sheets of
flatness}. These correspond to flat triangles. We call the region
a \textit{gauge invariant pyramid} \footnote{Note that not all of
the points inside the pyramid are gauge invariant because a
color-point (which represents the color condition of a vertex)
should also satisfy (\ref{even})}.

For a sandpile to be in a critical state, a fixed fraction of the
steps between sites must be at the critical value such that the
addition of one grain of sand will cause an avalanche of shifts of
grains.  The flat triangles play the same role in this model, they
are the triangles whose next evolution, by the addition or
subtraction of loops of flux, is likely to lead to
gauge-non-invariant configurations.  Hence in a critical state
there must be a fixed fraction of such flat triangles. We will see
that this expectation is satisfied when we find a set of rules
that generates scale invariant distributions of avalanches.

In Borissov and Gupta's model the vertices whose color-points are
inside the gauge invariant pyramid (and are not flat) always are
modified by the addition of positive colors. By adding a positive
color to one of the edges of such a vertex, its color-point goes
farther away from the origin of the color space. Roughly speaking,
in this situation the probability of finding the new color-point
on one of the sheets of flatness decreases. Therefore the
probability of producing a flat vertex by a non-flat vertex is not
high. In the simulation of the model it is clear that as time goes
on, only small number of avalanches happen and for this reason the
distribution of small size avalanches grows faster than larger
ones. Thus the model does not exhibit a power law distribution of
avalanches.\footnote{ The fraction of flat triangles is related to
the fact that those vertices which become flat initially remain on
the sheet of flatness and after a while no more color-points join
the sheets.}

\subsection{Random vertex models}

We now consider a different class of models, in which the evolution proceeds
by adding or subtracting color simultaneously on all edges incident to a single
vertex. We call these {\it random vertex models.}

The motivation for these models comes from looking at  Figure 3.
We see that if  we subtract color from all three  edges of a gauge
invariant vertex (like $B$), the new point will be closer to  one of
the sheets of flatness.
We then define a class of evolution rules  based on choosing nodes rather
than edges at random. In this class of rules, we will subtract the color $-2 $
from {\it all} three edges  incident on the chosen node.

The result can be to violate the gauge condition at that node
and/or at adjacent nodes. To recover gauge invariance we need to
define a propagation rule. One of the possible candidates is add
+2 to all the three incident edges at the GNI-vertex. We call this
simple rule the {\it triangle propagation rule} because it adds
three equal colors to a trivalent vertex.

To specify the rule we have to give an ordering to the nodes.  In
our simulations, we use a simple ordering, left to right and top
to bottom. We sweep the graph, checking the gauge invariance
condition (\ref{triangles},\ref{even}) at each node. When we find
a GNI we act with the triangle propagation rule, by adding $+2$ to
the colors on the edges incident to that node. When the checking
is done once for all of the vertices, the sweep is
 repeated because new GNI-vertices may have been
produced in the first sweep. We continue to repeat the propagation rule
until all vertices become gauge invariant.

For example, consider the following network:

\begin{center}
\includegraphics[width=180pt]{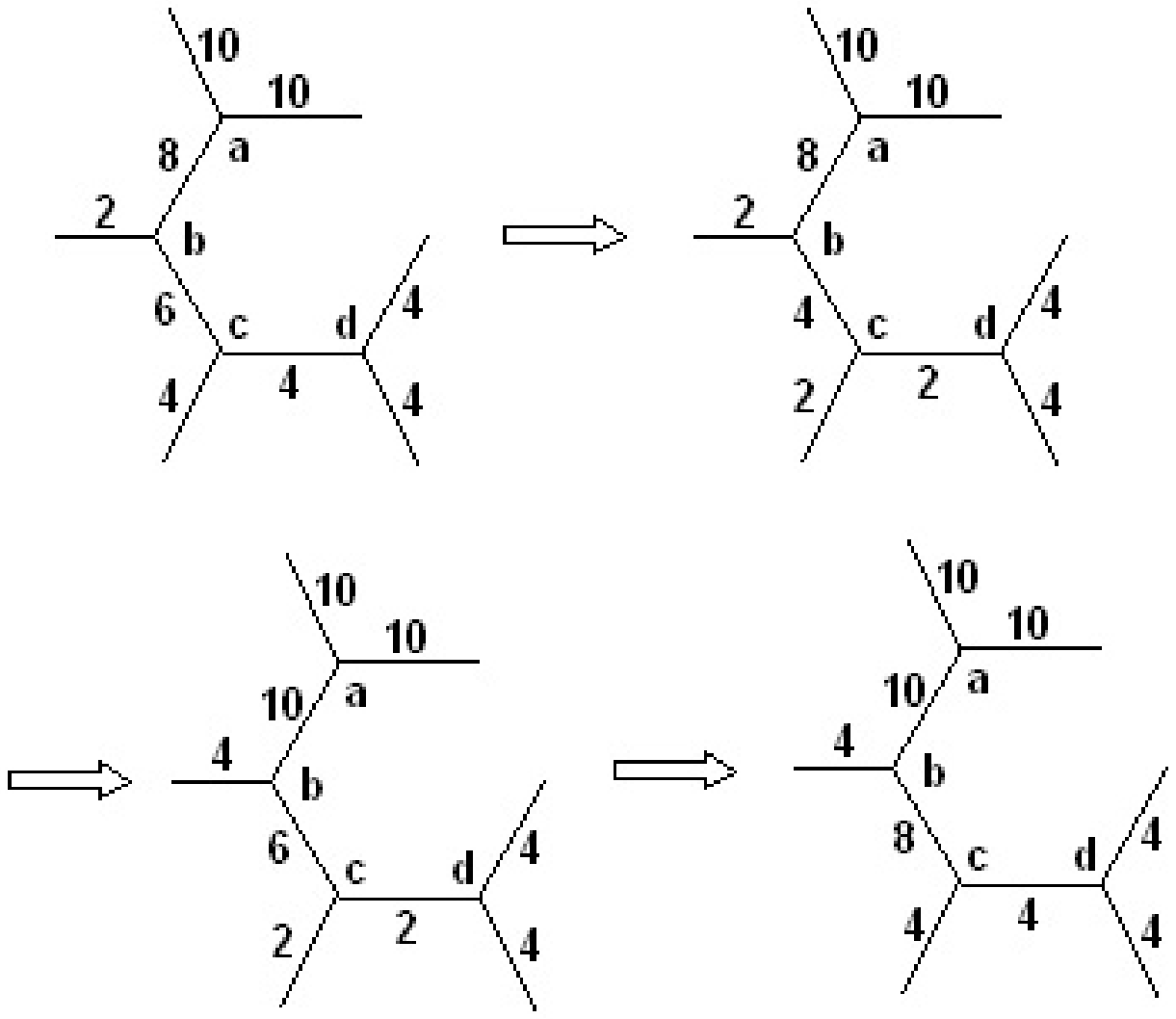}
\end{center}

The first diagram shows a simple network. In the second step, it
has been evolved by subtracting  -2  from each edge incident to
the vertex $c$. We then sweep the graph, from top to bottom and
from left to right. Vertex $c$ remains gauge invariant but $b$ is
not gauge invariant, so we act by adding $+2$ to each of its
incident edges.
 But this makes $c$ a GNI-vertex. In the
fourth step, +2  has been added to the edges incident to the vertex
$c$. Doing so, this new network becomes gauge invariant. The
number of steps in the avalanche in this evolution is $2$ because two vertices
were updated in order to make the network gauge invariant.

Let's summarize the random vertex class:

\begin{itemize}

\item Initialize the spin network by assigning random colors to its
edges, requiring only that  gauge invariance is satisfied at each node.
\item Choose a vertex at random.
\item Subtract a triangle of -2  from the three edges of the
initial vertex.
\item Check the gauge condition (\ref{triangles},\ref{even})
at all vertices, by sweeping through the nodes according to some
fixed rule. Fix each GNI-node by adding a triangle of +2 color to
each edge of its dual triangle.
\item Continue till the graph is again gauge invariant.  Count the
number of updated vertices.  This is the size of the avalanche.
\item Go back to the second step of the algorithm
and repeat.
\end{itemize}

\section{Results}

We now report on the simulation of the rule just described, which
did lead to scale invariant behavior.

We did the simulation for a 2d planar spin network with 570 edges
and 361 vertices. For the initial start we assigned random even
numbers between 10 to 30 to each edge, requiring only that the
graph be gauge invariant initially. Using the rules just
described, we evolved the spin network labels for ten million
steps. The result is shown in Figure 4.

\begin{figure}[h]
\label{loglog}
\begin{center}
\includegraphics[width=190pt]{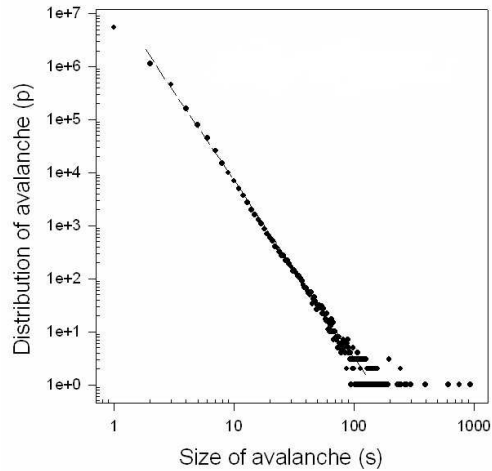}
\end{center}
\caption{The log-log plot of the distribution of avalanche in a 2d
planar spin network.}
\end{figure}

To completely define the evolution rule we mention:

\begin{itemize}
\item{}The nodes were always swept the same way, from left to right and up to down.
\item{}It happens often that the color on an edge is reduced to
$0$.  The rules act on such edges as on the others, with the one
exception that the $-2$ rule is never applied to a node when one
of its edges is $0$, as that would lead to an edge with a negative
color.  However the triangle propagation rule acts on triangles
with one or more edges zero as on other triangles. For example, a
triangle with colors $(0,18,18)$ is gauge invariant, and so is
skipped by the triangle propagation rule. But a triangle labelled
$(0,2,4)$ is fixed, by adding $+2$ to each edge. The result is
$(2,4,6)$, which is gauge invariant.
\end{itemize}

The distribution of the size of avalanches in a loglog scale
behaves, to a good extent, linearly. Thus the dynamics of the
triangle propagation rule on the spin network follows a power law
and exhibits abelian self-organized critical behavior. The
relation between the distribution of avalanche and the size of
avalanche is:
\begin{equation}
P(s) = s^{-3.3}
\end{equation} to a good accuracy.

In a SOC model usually both area and size of avalanches are
checked to behave power-law distributions. Area is the number of
sites involved in an avalanche, no matter how many times they
topple. In other word, the area is where the avalanche is taking
place, and usually for larger lattices one finds larger areas,
because it has a lattice dependent cut-off in its power-law
distribution. If this distribution instead of being power-law is
exponential, the avalanches do not expand in space and basically
it does not matter if one takes a small or large lattice, as long
as this is bigger than the maximum area that an exponential
distribution is likely to give in finite samples \cite{marco}.

To ensure this, we provide a typical plot of the distribution of
area of avalanches. Figure 5 indicates the power-law behavior of
area to a good accuracy in its distribution. Therefore, the
macroscopic emerging of avalanches in space can be observed during
this evolution of spin network. In other word, the avalanches do
not resemble of some local resonances in a few nodes.

\begin{figure}[h]
\begin{center}
\includegraphics[width=190pt]{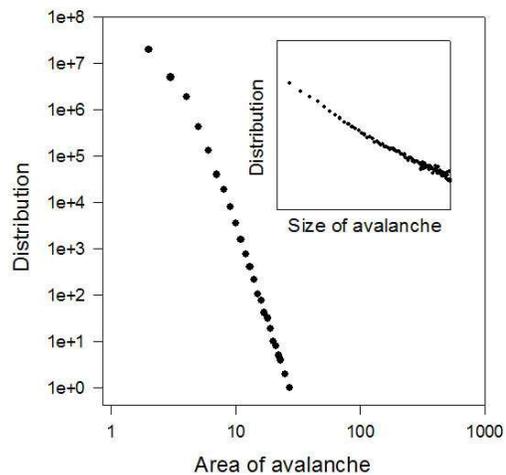}
\end{center}
\caption{A typical log-log plot of the distribution of the area of
avalanche and its corresponding log-log distribution of the size
of avalanche.}
\end{figure}

In each time step of the evolution, we recorded the average of
colors of the network and the fraction of the flat triangles,
which are the cause of avalanche. In Figure 6 we see that the
fraction of the number of flat vertices (or their dual triangles)
is maintained about 0.3 during the simulation, while in Figure 7
we see how the average color in the spin network fluctuate in
time.

\begin{figure}[h]
\begin{center}
\includegraphics[width=180pt]{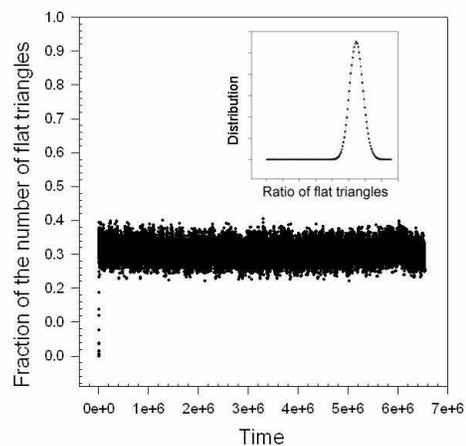}
\end{center}
\caption{The fraction of flat triangles in time.}
\end{figure}

As color is proportional to length, Figure 7 exhibits a universe,
described by our 2d planar spin network, expanding in time. The
model, this is an example\footnote{For another example see
\cite{bak 2002} and \cite{pacuzki 2002}.} of self-organization not
to an attractor state, but to an asymptote, on which the average
radius has a constant rate of inflation (expansion), is critical,
and exhibits avalanches of activity with power-law distributed
sizes. This example demonstrates that self-organized critical
behavior occurs in a larger class of systems than so far
considered: systems not driven to an attractive fixed point, but,
e.g., an asymptote, may also display self-organized criticality.

\begin{figure}[h]
\begin{center}
\includegraphics[width=190pt]{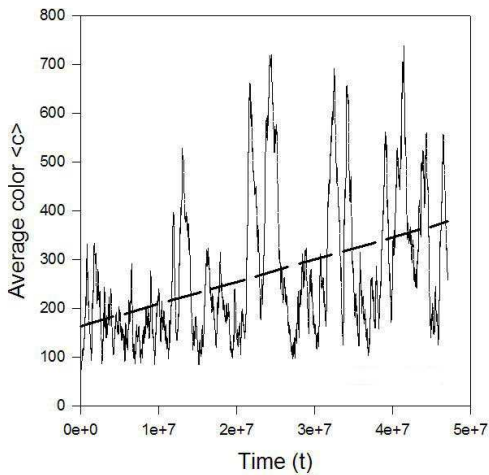}
\end{center}
\caption{The average color of the spin network in time in fifty
million iterations.}
\end{figure}

Finally, it is instructive to see how the evolution rule studied
here affects the dual geometry, expressed in terms of the
triangulation. In the Figure 8 we follow a piece of a dual spin
network, as it evolves.   We see the evolution is irregular in
both time and space.  Nevertheless, when averaged over large times
and distances, a scale invariant behavior emerges.

\begin{figure}[h]
\begin{center}
\includegraphics[width=150pt]{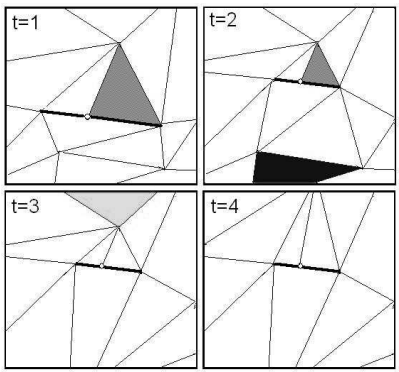}
\end{center}
\caption{A few steps in the evolution of a part of a dual
triangulation. The thick line (included a vertex in the middle)
represents a flat triangle. At t=2, a triangle of color -2 is
added to the shaded triangle, shrinking its sides.  The triangle
inequalities are violated on some neighboring triangles and these
are resolved by addition of a triangle of color +2 to them. At t=3
we then iterate the procedure subtracting 2 from the edges of the
black triangle and at t=4 we do the same to the bright gray one.}
\end{figure}

\section{Conclusions}

We have proposed a propagation rule for colors to evolve on a 2d planar open
spin network, which appears to  exhibit self-organized
critical behavior.

It appears that with a special choice of evolution rule,  the dynamics evolves the
system to a dynamical equilibrium state,
within which the behavior of the system appears to be scale invariant.

This work is a step in the investigation of the hypothesis that the emergence of our classical world from a
discrete quantum geometry is analogous to a self-organized critical process. Among the further steps are 1) the
study of models in which the underlying graphs themselves evolve by local rules, analogous to those studied here,
2) the study of other correlation functions, including those that would be interpreted as propagation amplitudes
for matter and gravitational degrees of freedom, and 3) an increase in the valence,
from three to four valent graphs, which is expected to correspond to the dynamics
of geometry in $3+1$ dimensions, and
4) the demonstration that self-organized critical phenomena
exists for quantum evolution and not just for ordinary statistical systems.

These are considerable challenges, towards which the present results must be seen as just a first step.

\section*{Acknowledgements}

We thank Maya Paczuski, Marco Baiesi, Artem Starodubtsev and
Fotini Markopoulou  for critical discussions about this project.
We are also grateful to Ali Tabei and Hoda Moazzen for
conversations on complex systems. We are grateful also to Poya
Haghnegahdar, Seth Major, Hamid Molavian and Tomasz Konopka for
their comments and advice on the manuscript.

\end{document}